\documentclass[review]{elsarticle}

\makeatletter
\def\ps@pprintTitle{%
 \let\@oddhead\@empty
 \let\@evenhead\@empty
 \def\@oddfoot{\centerline{\thepage}}%
 \let\@evenfoot\@oddfoot}
\makeatother

\usepackage{lineno,hyperref}
\modulolinenumbers[5]
\usepackage{soul,color}

\usepackage{amsmath}
\usepackage{amssymb}
\usepackage{graphicx}
\usepackage{subcaption}
\usepackage{lipsum}
\usepackage{float}
\usepackage{stackengine}

\usepackage[percent]{overpic}
\usepackage{enumerate}

\journal{Scripta Materialia}

\begin{document}

\begin{frontmatter}

\title{Effect of elastic anisotropy on phase separation in ternary alloys: A phase-field study}

\author[mymainaddress]{Sandeep Sugathan}
\ead{ms14resch01001@iith.ac.in}

\author[mymainaddress]{Saswata Bhattacharya\corref{mycorrespondingauthor}}
\cortext[mycorrespondingauthor]{Corresponding author}
\ead{saswata@iith.ac.in}

\address[mymainaddress]{Indian Institute of
Technology, Department of Materials Science and Metallurgical Engineering, Hyderabad, 502285, India}

\begin{abstract}
The precipitate shape, size and distribution are crucial factors which determine the properties 
of several technologically important alloys. Elastic interactions between the inclusions modify 
their morphology and align them along elastically favourable crystallographic
directions. Among the several factors contributing to
the elastic interaction energy between precipitating phases, anisotropy in elastic moduli is 
decisive in the emergence of modulated structures during phase separation in elastically 
coherent alloy systems. We employ a phase-field model incorporating elastic interaction
energy between the misfitting phases to study microstructural evolution in ternary 
three-phase alloy systems when the elastic moduli are anisotropic. 
The spatiotemporal evolution of the composition field variables is governed by solving a 
set of coupled Cahn-Hilliard equations numerically using a semi-implicit Fourier spectral technique.
We methodically vary the misfit strains, alloy chemistry and elastic anisotropy to investigate 
their influence on domain morphology during phase separation. The coherency strains between
the phases and alloy composition alter the coherent phase equilibria and decomposition pathways. The degree of anisotropy in elastic moduli modifies the elastic interaction
energy between the precipitates depending on the sign and magnitude of relative misfits, and
thus determines the shape and alignment of the inclusions in the microstructure. 
\end{abstract}


\end{frontmatter}


Elastic anisotropy significantly influences the microstructures in multicomponent, multiphase
alloys with coherent elastic misfit. The elastic interactions arising due to misfit between
the phases, anisotropy in elastic moduli, and elastic homogeneity produce changes in the
morphology and alignment of phases~\cite{khachaturyan2013theory,mura2013micromechanics}.
The studies on the influence of elastic anisotropy on morphological evolution in 
binary alloys was pioneered by Cahn in his seminal paper on spinodal decomposition in cubic 
crystals~\cite{cahn1962spinodal}. He showed that the coherency strain between coexisting phases 
and anisotropy in elastic moduli leads to decomposition occurring by sinusoidal composition
modulations along elastically preferred orientations depending upon the degree of elastic anisotropy..
He predicted the alignment of domains along elastically soft $\langle 100\rangle$ directions 
when $A_z>1$ and along $\langle 111\rangle$ directions when $A_z<1$. 

There is substantial experimental evidence on the modulated arrangement of precipitates in 
two-phase coherent alloys~\cite{ardell1966modulated,de1966microstructure,
butler1970structure,livak1971spinodally,
higgins1974precipitation,kubo1979spinodal,tyapkin88,tyapkin89,miyazaki1989shape,
sequeira1994shape,FAHRMANN19951007,inuzuka2007tem,kozakai2011substitution}. On investigating the modulated structures observed
in $Ni$-$Al$ alloys~\cite{ardell1966modulated}, Ardell and Nicholson
postulated that the alignment is caused by elastic interactions between precipitates. 
They also reported that the degree of alignment depends on volume fraction of the 
precipitates (supersaturation), misfit strains and elastic anisotropy.  

There have been several attempts at modeling the microstructural
evolution in elastically anisotropic two-phase alloy systems. Some of these studies focused on 
the equilibrium shape of misfitting particle in an elastically anisotropic 
medium~\cite{voorhees1992morphological,thompson1994equilibrium}. Computational
models were also employed to investigate coarsening in elastically anisotropic solids
with two or multiple coherent precipitates~\cite{su1996dynamics,su1996dynamics2,wang1992particle}. 
Several researchers simulated the development of modulated structures during both spinodal 
decomposition and nucleation and growth process in coherent binary 
alloys~\cite{nishimori1990pattern,wang1991strain,wang1993kinetics,Fratzl1996,hu2001phase}. 

With the above discussed motivation,
the purpose of our work is to the extend the understanding of the
role of elastic anisotropy in the emergence of modulated structures in binary alloys to ternary
three-phase alloys. In this regard, we develop a diffuse interface model to 
perform systematic simulations to 
investigate the combined effect of anisotropy in elastic moduli and the sign and degree misfit 
between coexisting phases on ternary phase separation. The presence 
of a third phase increases the complexity of elastic interactions and thus 
affects the direction of alignment of three phases differently.

We model a ternary substitutional alloy containing three atomic species A, B and C. 
The concentration of $i'$th species $c_i(\textbf{r},t)$ $(i=A,B,C)$ is a function of 
position \textbf{r} and time $t$, where $c_A+c_B+c_C=1$ according
to conservation condition.
We modify an existing diffuse-interface ternary Cahn-Hilliard model
~\cite{eyre1993systems,bhattacharyya2003study,ghosh2017particles} 
by introducing additional terms describing elastic interactions.

The chemical energy contribution to the total energy for the isotropic, compositionally 
inhomogeneous system expressed as a function of composition field variables is 
\begin{equation}
F = N_v\int_v\Big(\frac{1}{2}\sum_{i\neq j}\chi_{ij}c_ic_j+\sum_{i}c_i\ln{c_i} + \sum_{i}\kappa_i|\nabla c_i|^2\Big)dV,
\label{eq:energy}
\end{equation}

where $i,j=A,B,C$, $N_v$ is the number of molecules per unit volume (assumed to be independent of composition and position), $\chi_{AB}$, $\chi_{AC}$ and $\chi_{BC}$ are the pair-wise interaction parameters, and
$\kappa_i$ are the gradient energy coefficients associated with composition fields.

The elastic energy contribution to the total energy for the system
is expressed in reciprocal space using Khachaturyan's microelasticity
theory~\cite{khachaturyan2013theory}: 
\begin{equation}
F_{el}=\frac{1}{2}\sum_{p,q=0}^1\int\frac{d^3\textbf{k}}{(2\pi)^3}B_{pq}(\textbf{n})\tilde{\theta_p}(\textbf{k})\tilde{\theta_q}^*(\textbf{k}),
\label{eq: elastenk}
\end{equation}

where \textbf{k} denotes the Fourier wave vector, $\textbf{n}=\frac{\textbf{k}}{|\textbf{k}|}$ is the unit
vector in reciprocal space, $\tilde{\theta_p}(\textbf{k})$ represents the Fourier transform of 
$\theta_p(\textbf{r})$. $B_{pq}(\textbf{n})=\lambda_{ijkl}\epsilon_{ij}^{(p)}\epsilon_{kl}^{(q)}
-n_i\hat{\sigma_{ij}}^{(p)}\omega_{jk}(\textbf{n})\hat{\sigma_{kl}}^{(q)}n_l$
is the elastic interaction energy between $\theta_p$ and $\theta_q$
and $\omega_{il}^{-1}(\textbf{n})=\lambda_{ijkl}n_jn_k$ is the inverse Green tensor.
$\tilde{\theta^*}$ denotes the complex conjugate of $\tilde{\theta}$. 

The temporal evolution of the conserved field variables $c_B(\textbf{r},t)$
and $c_C(\textbf{r},t)$ ($\because c_A+c_B+c_C=1$) is governed by a set of
Cahn-Hilliard equations:
\begin{equation}
\begin{split}
\frac{\partial c_B}{\partial t}=M_{BB} & \nabla^2\Bigg(g_B-2\kappa_{BB}\nabla^2c_B-2\kappa_{BC}\nabla^2c_C+\frac{\delta F_{el}}{\delta c_B}\Bigg)+ \\ M_{BC} & \nabla^2\Bigg(g_C-2\kappa_{BC}\nabla^2c_B-2\kappa_{CC}\nabla^2c_C+\frac{\delta F_{el}}{\delta c_C}\Bigg),
\end{split}
\label{eq: dCbdt}
\end{equation}
\begin{equation}
\begin{split}
\frac{\partial c_C}{\partial t}=M_{BC} & \nabla^2\Bigg(g_B-2\kappa_{BB}\nabla^2c_B-2\kappa_{BC}\nabla^2c_C+\frac{\delta F_{el}}{\delta c_B}\Bigg)+ \\ M_{CC} & \nabla^2\Bigg(g_C-2\kappa_{BC}\nabla^2c_B-2\kappa_{CC}\nabla^2c_C+\frac{\delta F_{el}}{\delta c_C}\Bigg),
\end{split}
\label{eq: dCcdt}
\end{equation}

where $\frac{\delta F_{el}}{\delta c_B}$ and $\frac{\delta F_{el}}{\delta c_C}$
are variational derivatives of elastic energy with respect to composition
fields, $g_B$ and $g_C$ are bulk driving forces with respect to
composition fields, $\kappa_{BB}=\kappa_A+\kappa_B, 
\kappa_{BC}=\kappa_{CB}=\kappa_A, \kappa_{CC}=\kappa_A+\kappa_C$.
$M_{BB}$, $M_{BC}$ and $M_{CC}$ are Onsager kinetic coefficients defined as follows \cite{allnatt_lidiard_1993}:
\begin{equation}
M_{BB}=(1-c_B)^2M_{B}+c_B^2(M_A+M_C),
\nonumber
\end{equation}
\begin{equation}
M_{CC}=(1-c_C)^2M_{C}+c_C^2(M_A+M_B),
\nonumber
\end{equation}
\begin{equation}
M_{BC}=M_{CB}=c_Bc_CM_A-c_B(1-c_C)M_{C}-c_C(1-c_B)M_{B},
\end{equation}

where $M_A$, $M_B$, and $M_C$ are the mobilities of species $A$, $B$, and $C$, respectively.

We use a semi implicit Fourier spectral method \cite{chen1998applications,PhysRevE.60.3564} to 
solve the coupled evolution equations 
(Eqns.~\ref{eq: dCbdt} and~\ref{eq: dCcdt}).
All parameters used in the simulation are rendered non-dimensional using
characteristic length, time and energy values. 
We perform two dimensional simulations of microstructural evolution
on a square grid of size $1024\times 1024$
with dimensionless grid spacing $\Delta x =\Delta y = 1$ employing 
periodic boundary conditions. 

The simulations start with a homogeneous alloy of a prescribed composition. 
A sustained Gaussian
noise of strength $0.5\%$ is added for the initial 10000 steps to mimic thermal fluctuations. 
We choose a non-dimensional time step $\Delta t=0.05$ for
evolving the phase-field variables. We choose the bulk free energy coefficients and the gradient energy coefficients 
to be equal ($\chi_{AB}=\chi_{AC}=\chi_{BC}=3.5$, $\kappa_A=\kappa_B=\kappa_C=4$)
to ensure a symmetric ternary miscibility gap with equal interfacial energies 
between the equilibrium phases 
($\Gamma_{\alpha\beta}=\Gamma_{\alpha\gamma}=\Gamma_{\beta\gamma}$). 

We introduce anisotropy in elastic moduli by fixing the Zener anisotropy
ratio ($A_z=3$) and systematically vary the alloy composition and 
the sign and magnitude of relative misfit between coexisting phases
to understand the role of anisotropy in elastic moduli on microstructural
evolution during ternary phase separation.
We have chosen six different alloy systems categorized
depending on their chemical composition ($c_B, c_C$) and misfit 
strains ($\epsilon_{\alpha\beta},\epsilon_{\alpha\gamma}$) as listed in
Table~\ref{cases}.      
\begin{table}[ht!]
\caption{Alloy systems used for simulations  
(prescribed alloy composition and magnitude and sign of misfit strains)}
\label{cases}
\begin{center}
\begin{tabular}{ |c|c|c|c|c|c| } 
 \hline
 System & $c_B$ & $c_C$  & $\epsilon_{\alpha\beta}$ & $\epsilon_{\alpha\gamma}$\\
 \hline 
      $X_1$ & 0.15 & 0.15 & 0.01000 &  0.01000 \\
      $X_2$ & 0.15 & 0.15 & 0.01000 & -0.01000 \\
      $X_3$ & 0.15 & 0.15 & 0.01265 & -0.00632 \\
      $Y_1$ & 0.40 & 0.40 & 0.01000 &  0.01000 \\
      $Y_2$ & 0.40 & 0.40 & 0.01000 & -0.01000 \\
      $Y_3$ & 0.40 & 0.40 & 0.01265 &  0.00632 \\
 \hline
\end{tabular}
\end{center}
\end{table}

We follow a particular naming convention to specify an alloy system:
an uppercase letter denotes the composition of the alloy and an integer subscript represents the state of misfit strain.
Integer $`1'$ represents the case where the lattice expansion coefficients, 
$\epsilon_{\alpha\beta}$ and $\epsilon_{\alpha\gamma}$, associated with 
concentrations of B and C, respectively, have equal magnitude and same sign.
On the other hand, integer $`2'$ represents the case where 
$\epsilon_{\alpha\beta}=-\epsilon_{\alpha\gamma}$, and integer $`3'$ represents the case where 
$|\epsilon_{\alpha\beta}|\neq|\epsilon_{\alpha\gamma}|$.

The microstructures of the alloys are represented using an RGB 
(red-green-blue) color map indicating the local compositions of
components $A$, $B$, and $C$. As per the color map, blue shade 
represents $A-$rich $\alpha$ phase, green represents $B-$rich $\beta$ phase,
and red represents $C-$rich $\gamma$ phase. We specify the interfaces by
the linear combination of three terminal colors depending upon the compositions.

\begin{figure}[ht!]
\centering
\begin{subfigure}[b]{.3\textwidth}
  \centering
  \def\big{\includegraphics[height=3cm]{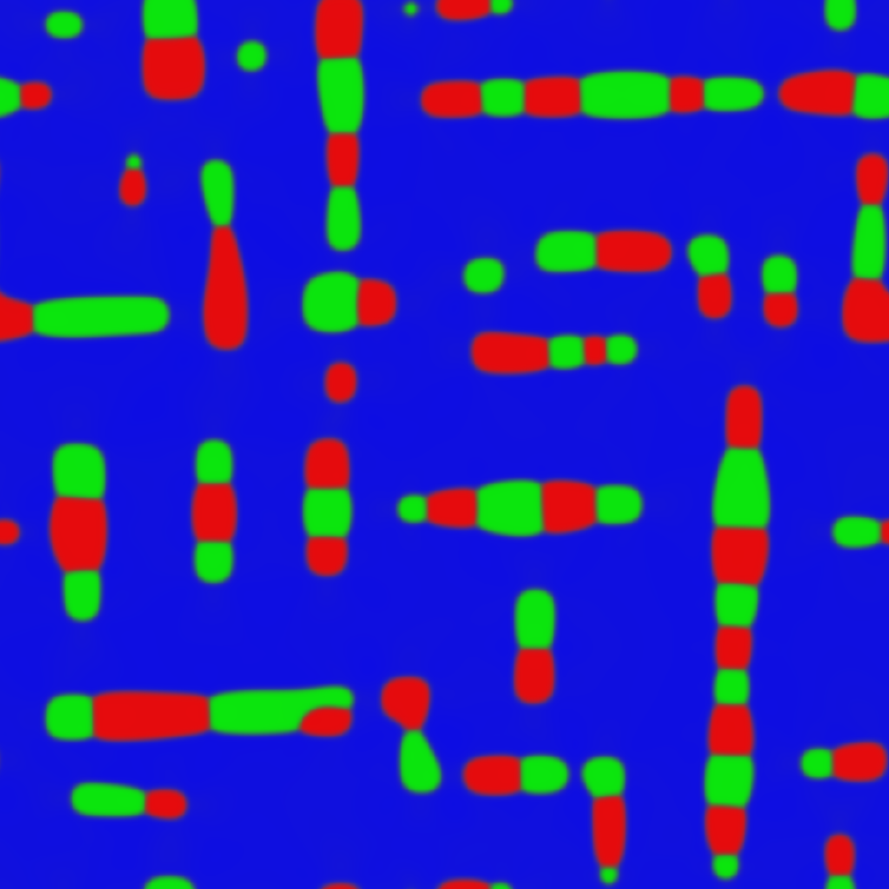}}
  \def\little{\includegraphics[height=1.2cm]{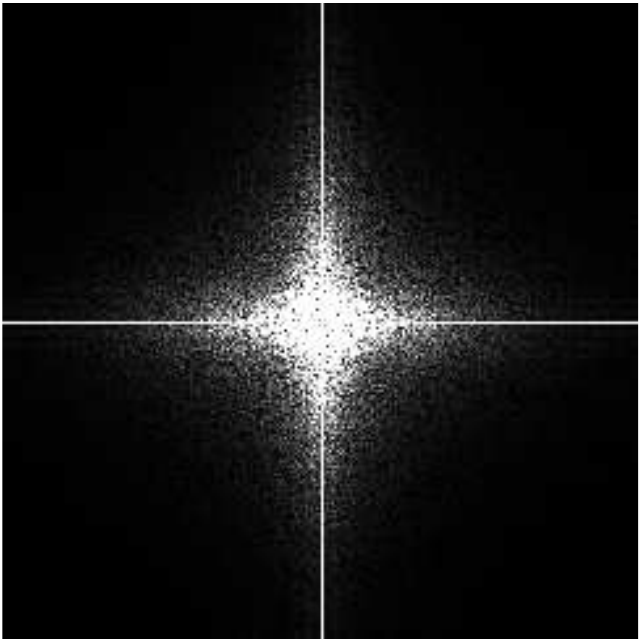}}
  \def\stackalignment{r}
  \bottominset{\little}{\big}{0pt}{0pt}
  \caption{$X_1$}
  \label{ms_x1}
\end{subfigure}%
\begin{subfigure}[b]{.3\textwidth}
  \centering
  \def\big{\includegraphics[height=3cm]{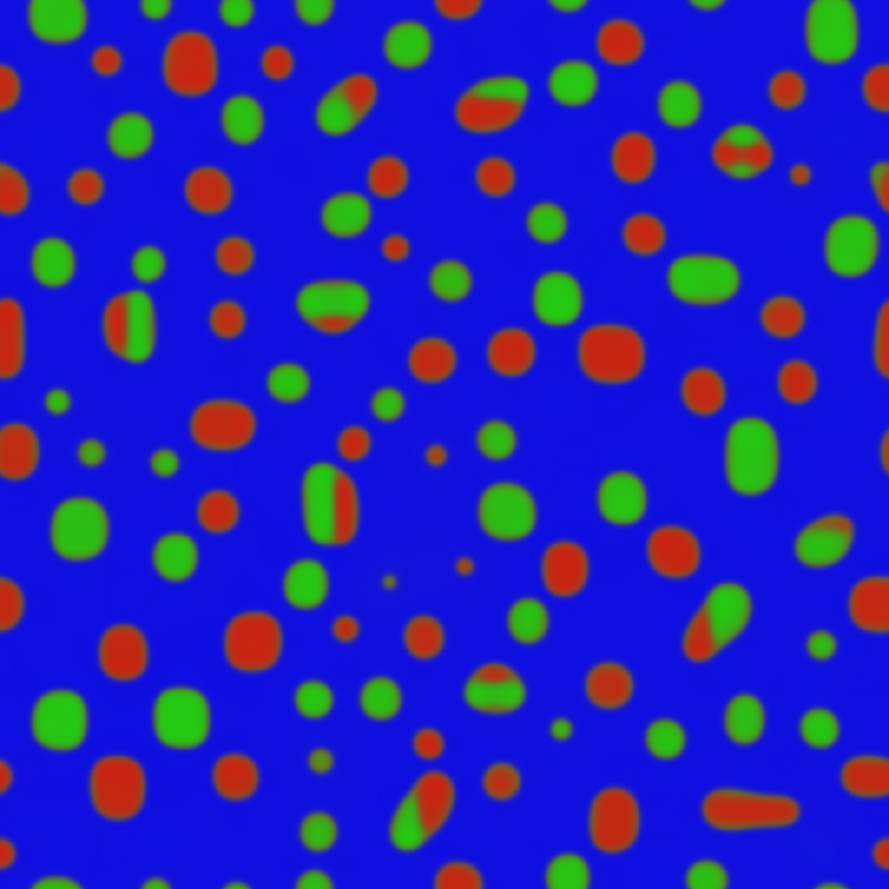}}
  \def\little{\includegraphics[height=1.2cm]{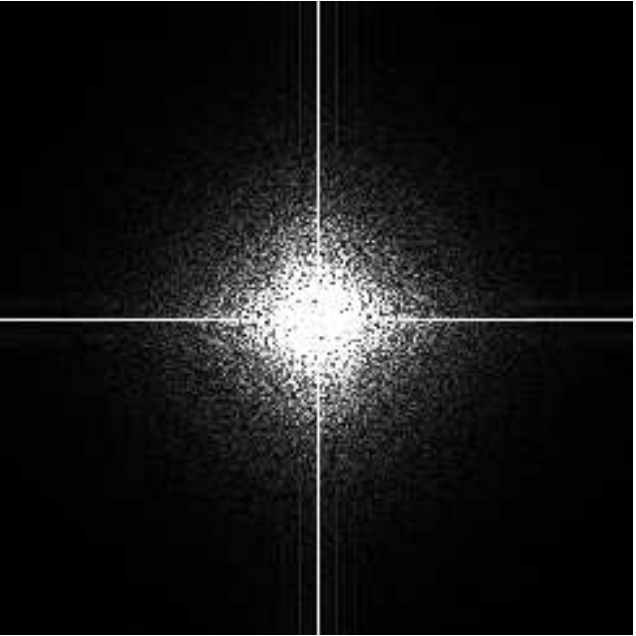}}
  \def\stackalignment{r}
  \bottominset{\little}{\big}{0pt}{0pt}  
  \caption{$X_2$}
  \label{ms_x2}
\end{subfigure}
\begin{subfigure}[b]{.3\textwidth}
  \centering
  \def\big{\includegraphics[height=3cm]{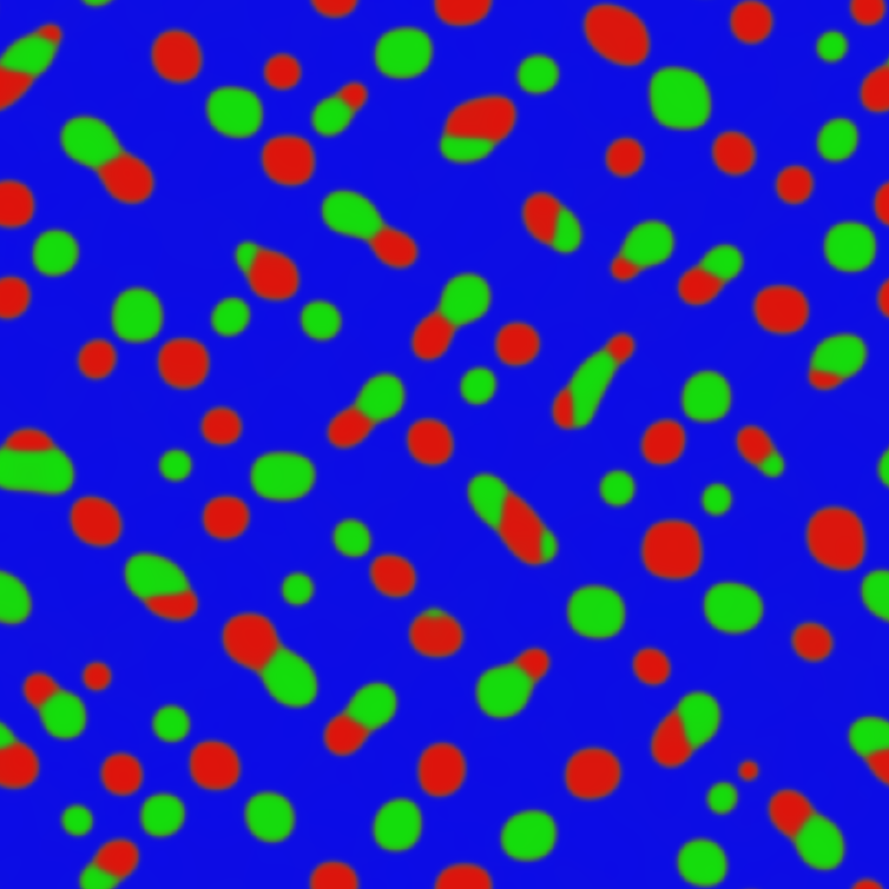}}
  \def\little{\includegraphics[height=1.2cm]{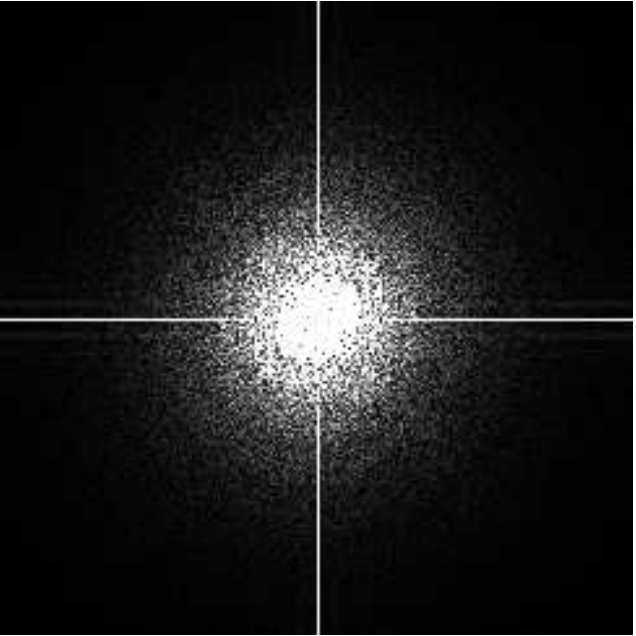}}
  \def\stackalignment{r}
  \bottominset{\little}{\big}{0pt}{0pt}  
  \caption{$X_3$}
  \label{ms_x3}
\end{subfigure} 
\\
\begin{subfigure}[b]{.3\textwidth}
  \centering
  \def\big{\includegraphics[height=3cm]{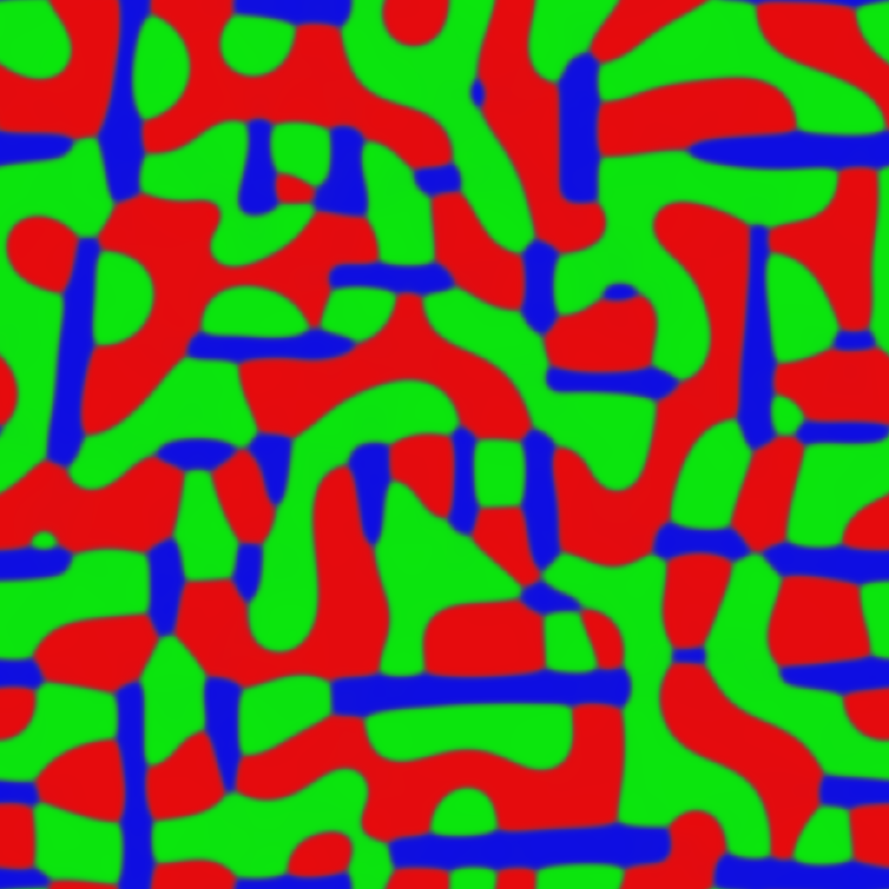}}
  \def\little{\includegraphics[height=1.2cm]{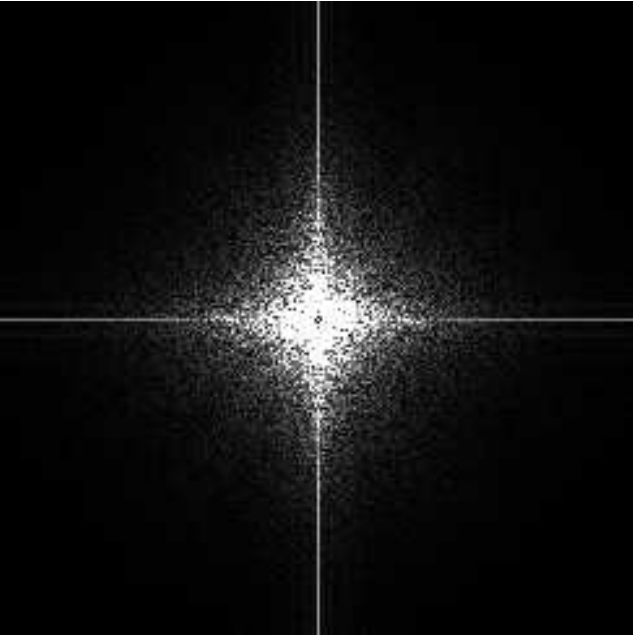}}
  \def\stackalignment{r}
  \bottominset{\little}{\big}{0pt}{0pt}
  \caption{$Y_1$}
  \label{ms_y1}
\end{subfigure}%
\begin{subfigure}[b]{.3\textwidth}
  \centering
  \def\big{\includegraphics[height=3cm]{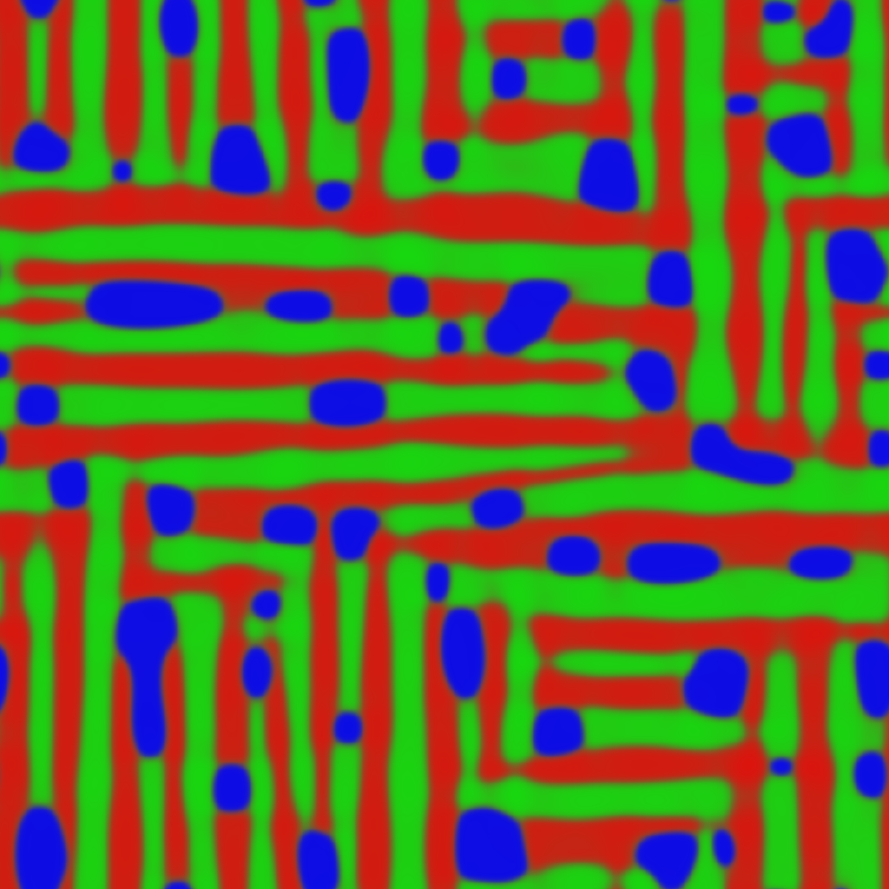}}
  \def\little{\includegraphics[height=1.2cm]{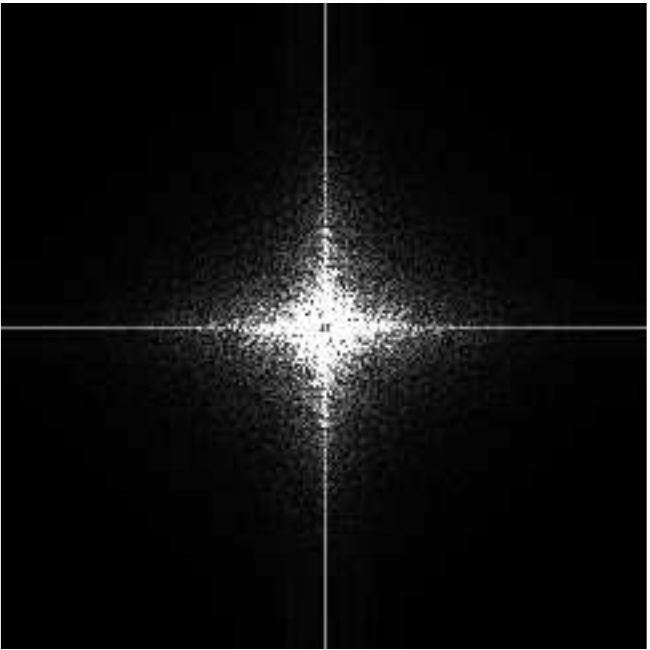}}
  \def\stackalignment{r}
  \bottominset{\little}{\big}{0pt}{0pt}  
  \caption{$Y_2$}
  \label{ms_y2}
\end{subfigure}
\begin{subfigure}[b]{.3\textwidth}
  \centering
  \def\big{\includegraphics[height=3cm]{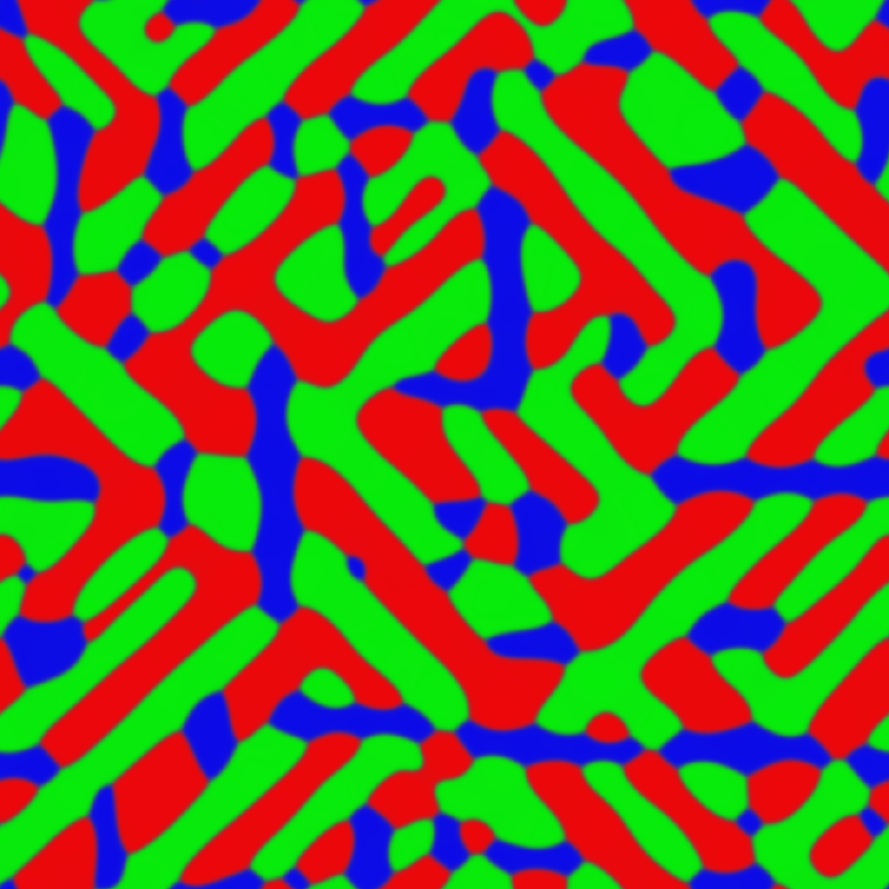}}
  \def\little{\includegraphics[height=1.2cm]{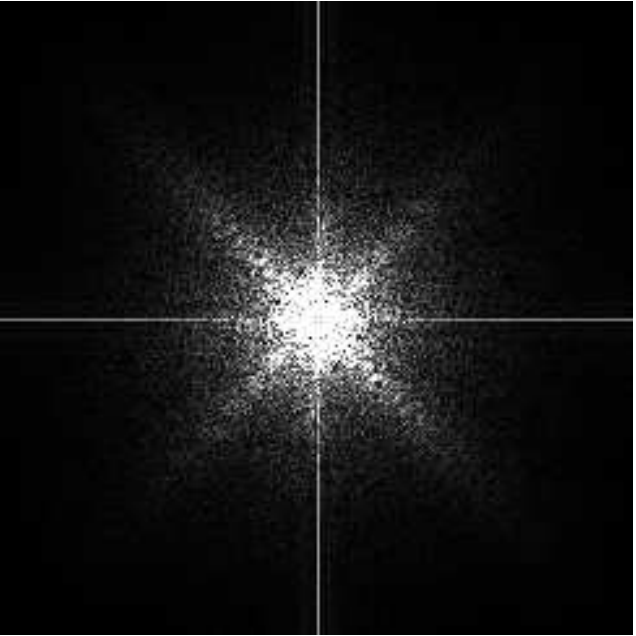}}
  \def\stackalignment{r}
  \bottominset{\little}{\big}{0pt}{0pt}  
  \caption{$Y_3$}
  \label{ms_y3}
\end{subfigure} 
\caption{Time snapshots of microstructure of alloy systems $X$, and $Y$ at $t=20000$. 
The power spectrum corresponding to each microstructure is given in the inset of respective figure.}
\label{mstruct}
\end{figure}

The time snapshots of the microstructures of $A-$rich $X$ alloys and $A-$poor $Y$ alloys 
and power spectra corresponding to the microstructures are shown in Fig.~\ref{mstruct}. 
Power spectrum is the measure of the power content of a signal as a function of it’s frequency.
The spectrum of the microstructure will contain information about morphological anisotropy.
The power spectrum analysis of a simulated microstructure is analogous to the small angle 
X-ray scattering (SAXS) data obtained from an experimental microstructure.

In $A-$rich  $X$ alloys, the initial phase separation
takes place along the median perpendicular to $BC$ line leading to the formation
of $A-$poor domains in an $A-$rich matrix. The `$BC$'-rich
domains undergo secondary decomposition forming $\beta$ and $\gamma$ phases.
As per the Zener ratio, $\langle 10\rangle$ directions are the elastically
soft directions for these alloys.  

Since $\alpha$ is associated with non-zero misfit, primary decomposition into
$A-$poor and $A-$rich domains is sluggish in $X_1$.
But the secondary decomposition of the $A-$poor domains to $\beta$ and
$\gamma$ phases takes place quickly because of the absence of any misfit between
the two phases ($\epsilon_{\beta\gamma}=0$). 
The power spectrum of the microstructure has its peaks at 
$\langle 10\rangle$ directions indicating alignment
of $\beta$ and $\gamma$ domains along these elastically favorable directions (Fig.~\ref{ms_x1}).
This orientational preference is attributed to the equal and positive misfit of these phases with 
matrix phase $\alpha$. Since the $\beta\gamma$ interface is strain-free
($\epsilon_{\beta\gamma}=0$), the system prefers more $\beta\gamma$ interfaces
for the minimization of elastic energy. Thus, we observe chains comprising alternating
beads of $\beta$ and $\gamma$ aligned along $\langle 10\rangle$ directions in this alloy. 

The initial phase separation begins quickly in $X_2$ and $X_3$
since the misfits of $\beta$ and $\gamma$ phases with the matrix phase $\alpha$
are opposite in sign. But the secondary decomposition of the $A-$poor
domains is delayed because of high misfit between $\beta$ and
$\gamma$. The strain is highest in alloy $X_2$ ($\epsilon_{\beta\gamma}=0.02$). 
Therefore both domains appear much later. Since the $\beta\gamma$ interfaces 
have large misfit in both alloys, the domains remain isolated in the matrix.

In alloy $X_2$, the domains show a weak inclination towards $\langle 10\rangle$ 
directions since misfits are equal in magnitude, but opposite in sign 
($\epsilon_{\alpha\beta}=-\epsilon_{\alpha\gamma}=1.0$). Therefore, the power spectrum 
is diamond shaped indicating the weak alignment of isolated $\beta$ and $\gamma$ particles
along $\langle 10\rangle$ directions (Fig.~\ref{ms_x2}).
In $X_3$, $\beta$ and $\gamma$ have unequal misfits of opposite
sign with the matrix $\alpha$ phase. This leads to shape changes of the isolated 
domains from cuboidal to rhombohedral/diamond-like. 
The power spectrum is diffused and has circular
symmetry (Fig.~\ref{ms_x3}). This is because of the weak alignment of the isolated
domains along $\langle 10\rangle$ and interconnected domains along $\langle 11\rangle$. 

The $A-$poor $Y$ alloys phase separate initially to $B-$rich and $C-$rich domains and
$\alpha$ phase appears later during secondary decomposition. These alloys also have 
Zener anisotropy ratio $A_z=3$ and hence the $\langle 10\rangle$ 
directions are elastically soft. 

In $Y_1$ and $Y_3$, the initial decomposition to majority
phases $\beta$ and $\gamma$ happens quickly due to
the low coherency strains between those phases. 
$\epsilon_{\beta\gamma}=0$ for $Y_1$ alloy and 
$\epsilon_{\beta\gamma}=0.006$ for $Y_3$.
Since $\epsilon_{\beta\gamma}=0$, the 
$\beta\gamma$ interfaces are curved in $Y_1$ and the domains
do not show any specific directional alignment. However, 
in $Y_3$, since the $\beta$ and $\gamma$ phases have unequal 
misfits of same sign with the third phase $\alpha$,
both the $\beta$ and $\gamma$ domains align along $\langle 11\rangle$ 
directions. 

In both the alloys, $\alpha$ phase separates out eventually,
forming elongated domains along elastically soft $\langle 10\rangle$ 
directions. Due to this orientational preference shown by the $\alpha$ domains, 
the peaks for the power spectrum of $Y_1$ are at $\langle 10\rangle$ directions (Fig.~\ref{ms_y1}). 
Whereas, the power spectrum of $Y_3$ peaks at both $\langle 11\rangle$ and $\langle 10\rangle$ directions
owing to the strong alignment of minority phase $\alpha$ along $\langle 10\rangle$ and 
majority phases $\beta$ and $\gamma$ along $\langle 11\rangle$ directions (Fig.~\ref{ms_y3}).

The initial phase separation of $\beta$ and $\gamma$ domains is delayed in $Y_2$
since the relative misfit between them is large ($\epsilon_{\beta\gamma}=0.02$). 
Subsequently, after the formation of $\beta$ and $\gamma$, isolated domains of $\alpha$ appear by secondary spinodal decomposition. 
The $\alpha$ domains exhibit weak alignment along the elastically preferred $\langle 10\rangle$ 
directions. The $\beta$ and $\gamma$ phases also align along the same directions. 
Therefore, we observe only $\langle 10\rangle$ peaks in the power spectrum (Fig.~\ref{ms_y2}).  
 
Since interfacial energies between the coexisting phases are assumed to be
isotropic, the change in orientation of the domains as a function anisotropy in elastic moduli
can be explained from the corresponding polar plot of elastic interaction
energies as function of orientation with respect to the crystal axes.
The elastic interaction energy $B_{pq}(\textbf{n})$
between two domains $p$ and $q$ is given by the expression:
\begin{equation}
B_{pq}(\textbf{n})=\lambda_{ijkl}\epsilon_{ij}^{(p)}\epsilon_{kl}^{(q)}
-n_i\hat{\sigma_{ij}}^{(p)}\omega_{jk}(\textbf{n})\hat{\sigma_{kl}}^{(q)}n_l,
\end{equation}

where $\lambda_{ijkl}$ is the elastic stiffness tensor,
$\omega_{il}^{-1}(\textbf{n})=\lambda_{ijkl}n_jn_k$ is the normalized inverse Green tensor,
$\epsilon_{ij}^{(p)}$
is the eigenstrain field and $\hat{\sigma_{ij}}^{(p)}$ is the stress field associated
with respective domains. The inner envelop of the polar plots of $B_{pq}(\textbf{n})$ 
predicts the equilibrium shapes of domains that minimize the interaction energy.
\begin{figure}[ht!]
\centering
\begin{subfigure}[b]{.33\textwidth}
  \centering
  \includegraphics[width=0.8\linewidth]{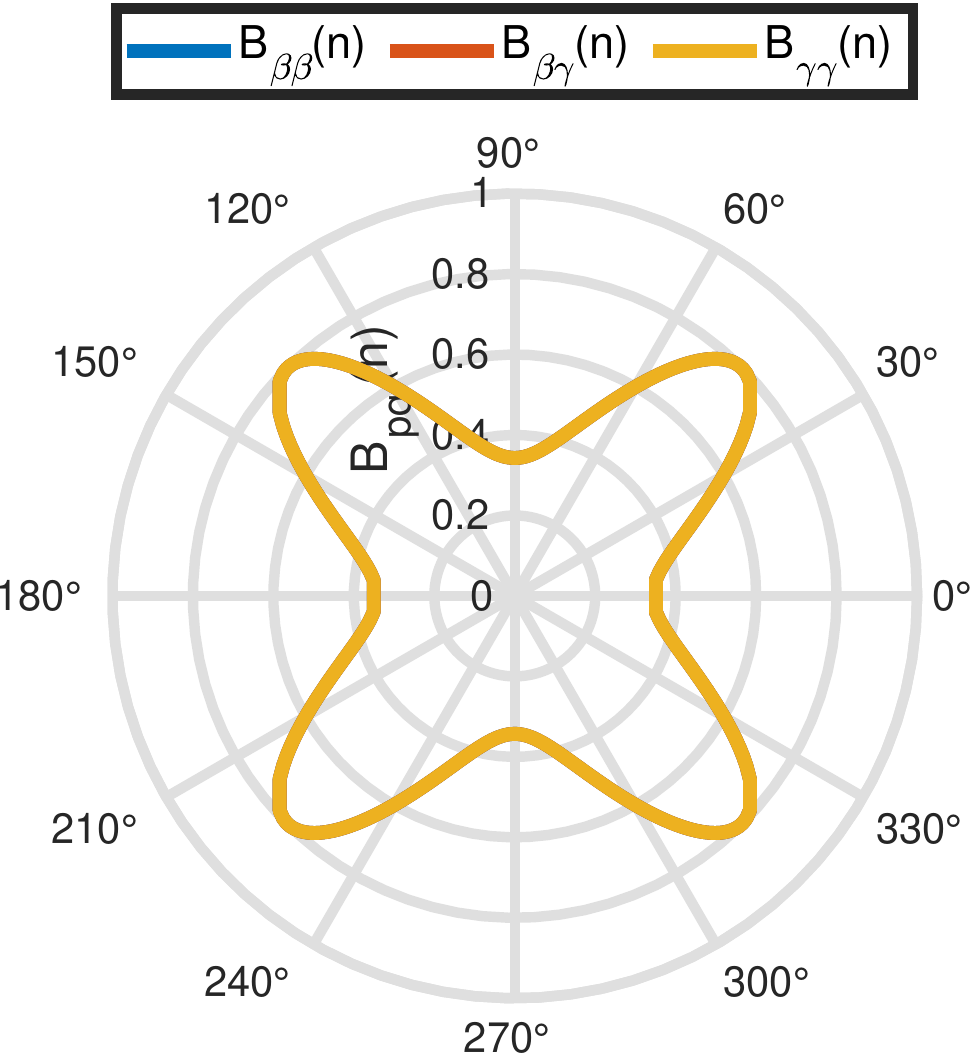}
  \caption{}
  \label{pol_x1}
\end{subfigure}%
\begin{subfigure}[b]{.33\textwidth}
  \centering
  \includegraphics[width=0.8\linewidth]{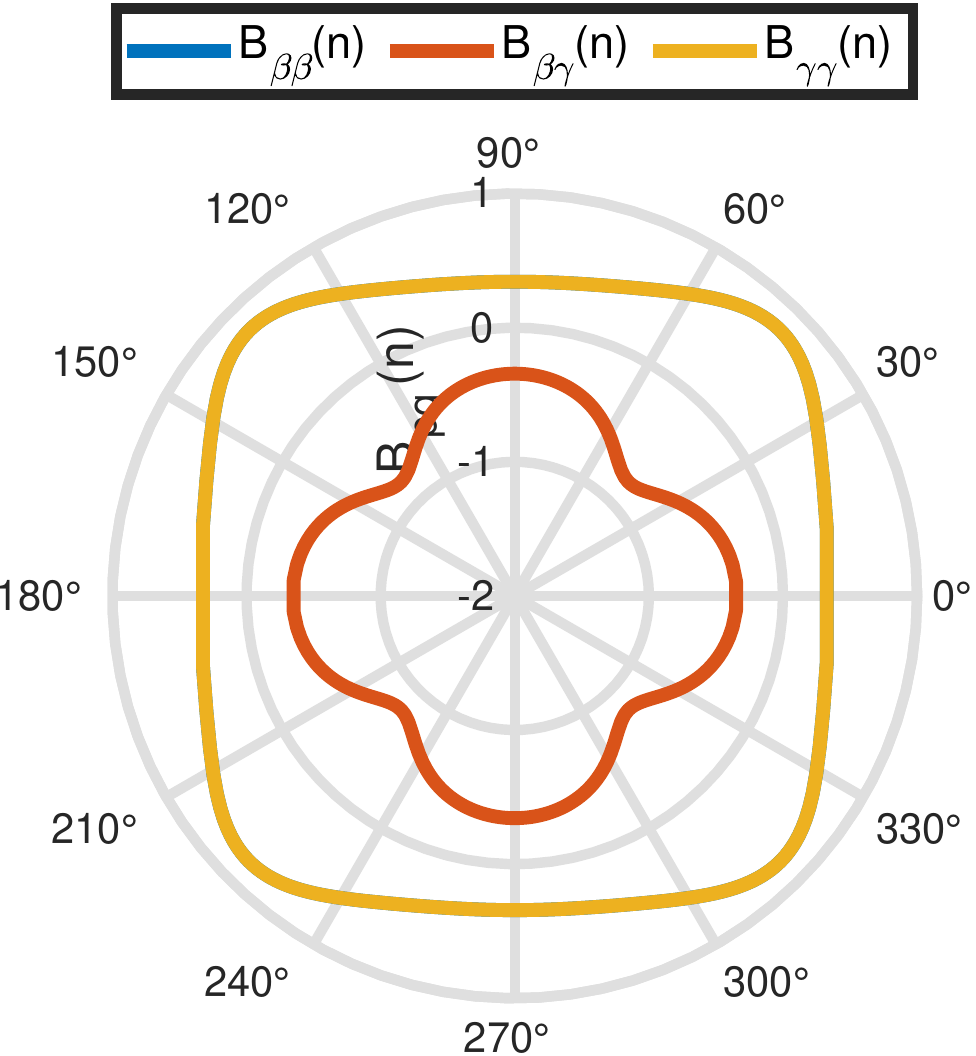}
  \caption{}
  \label{pol_x2}
\end{subfigure}
\begin{subfigure}[b]{.33\textwidth}
  \centering
  \includegraphics[width=0.8\linewidth]{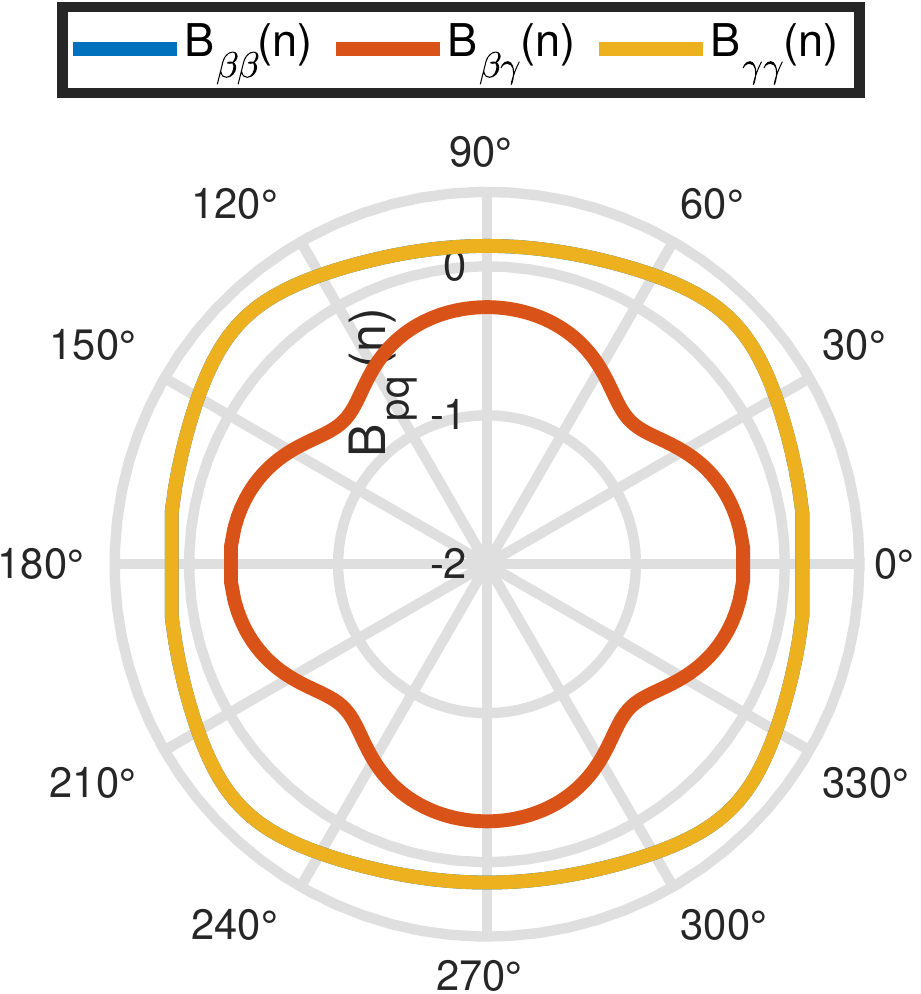}
  \caption{}
  \label{pol_x3}
\end{subfigure} 
\caption{Polar plots of elastic interaction energies $B_{pq}(\textbf{n})$
for alloys (a)$X_1$, (a)$X_2$ and (c)$X_3$. Each polar plot contains the
interaction energy between $\beta$ domains $B_{\beta\beta}(\textbf{n})$, 
interaction energy between $\gamma$
domains $B_{\gamma\gamma}(\textbf{n})$ and 
interaction energy between $\beta$ and $\gamma$ domains $B_{\beta\gamma}(\textbf{n})$.}
\label{fig:elinten_x123}
\end{figure}

Since $X_1$, $X_2$ and $X_3$ alloys have $\beta$ and $\gamma$
domains distributed in $\alpha$ matrix, we explain the morphological patterns 
according to $B_{\beta\beta}(\textbf{n})$, 
$B_{\gamma\gamma}(\textbf{n})$ and
$B_{\beta\gamma}(\textbf{n})$, where $B_{\beta\beta}(\textbf{n})$ denotes 
interaction energy between $\beta$ domains, $B_{\gamma\gamma}(\textbf{n})$ denotes
interaction energy between $\gamma$ domains, and $B_{\beta\gamma}(\textbf{n})$ 
denotes interaction energy
between $\beta$ and $\gamma$ domains. As shown in Fig.~\ref{pol_x1},
the three interaction energies are equal for alloy $X_1$ and they show minima
when $\textbf{n}$ corresponds to $\langle 10\rangle$ directions. This explains the arrangement
of the domains along $\langle 10\rangle$ directions in $X_1$.

For $X_2$ and $X_3$ alloys, $B_{\beta\gamma}(\textbf{n})$ is not equal to 
$B_{\beta\beta}(\textbf{n})$ and $B_{\gamma\gamma}(\textbf{n})$ due to the changes in
sign and magnitude of the misfits. 
The interactions between $B_{\beta\beta}(\textbf{n})$, 
$B_{\gamma\gamma}(\textbf{n})$ and
$B_{\beta\gamma}(\textbf{n})$ change the elastically soft directions for alloys
$X_2$ and $X_3$. In $X_2$, $B_{\beta\beta}(\textbf{n})$ and
$B_{\gamma\gamma}(\textbf{n})$ show minima along $\langle 10\rangle$ when $B_{\beta\gamma}(\textbf{n})$
shows minima along $\langle 11\rangle$ (Fig.~\ref{pol_x2}). Thus isolated $\beta$ and $\gamma$ particles align
along $\langle 10\rangle$ and appear rectangular. However, there is a weak alignment of 
alternate $\beta$ and $\gamma$ particles along $\langle 11\rangle$. In $X_3$,
the interactions are similar to $X_2$ (Fig.~\ref{pol_x3}). However, the alignment of
alternate $\beta$ and $\gamma$ domains along $\langle 11\rangle$ seems stronger.
\begin{figure}[ht!]
\centering
\begin{subfigure}[b]{.33\textwidth}
  \centering
  \includegraphics[width=0.8\linewidth]{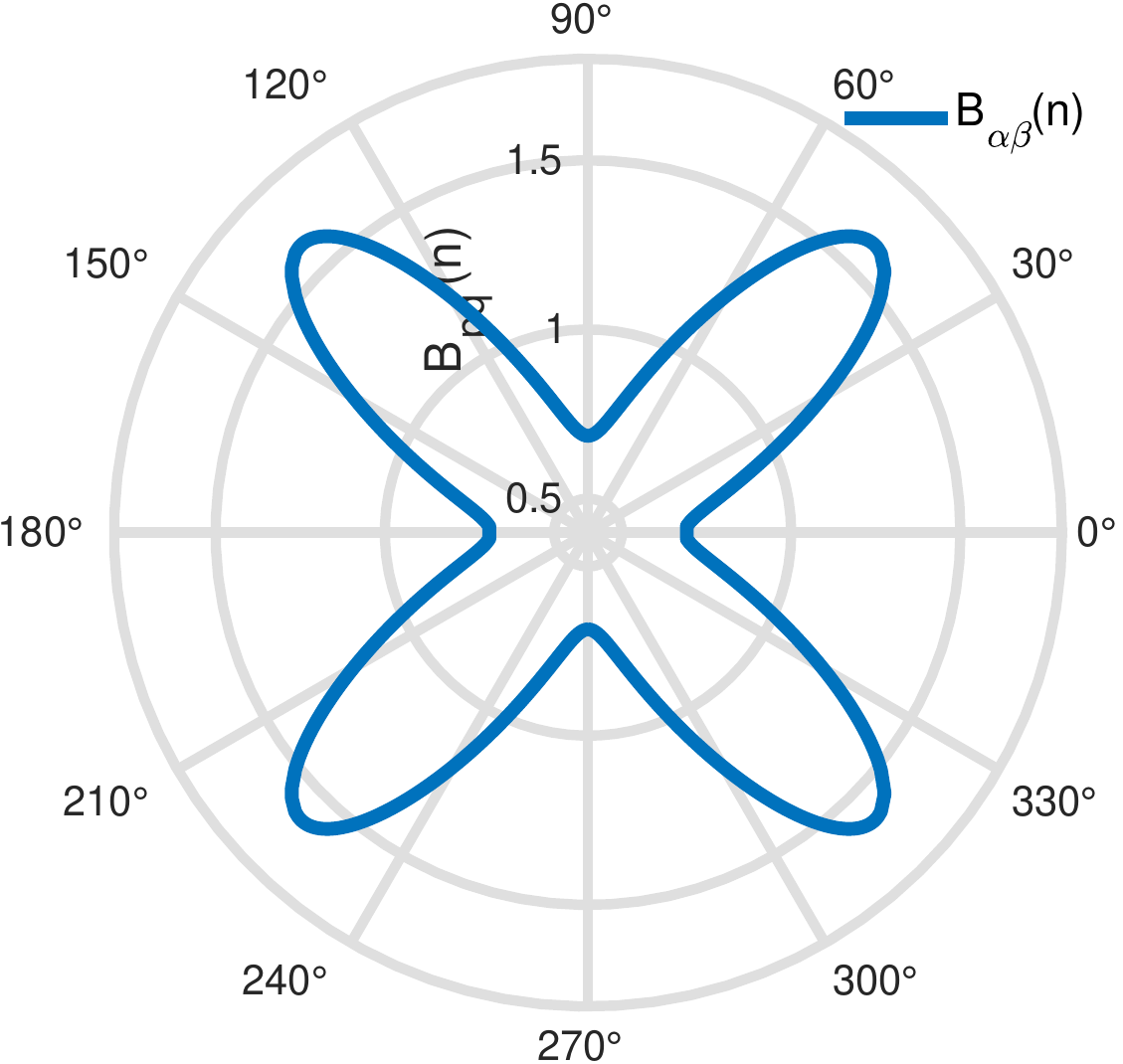}
  \caption{}
  \label{pol_y2}
\end{subfigure}
\begin{subfigure}[b]{.33\textwidth}
  \centering
  \includegraphics[width=0.8\linewidth]{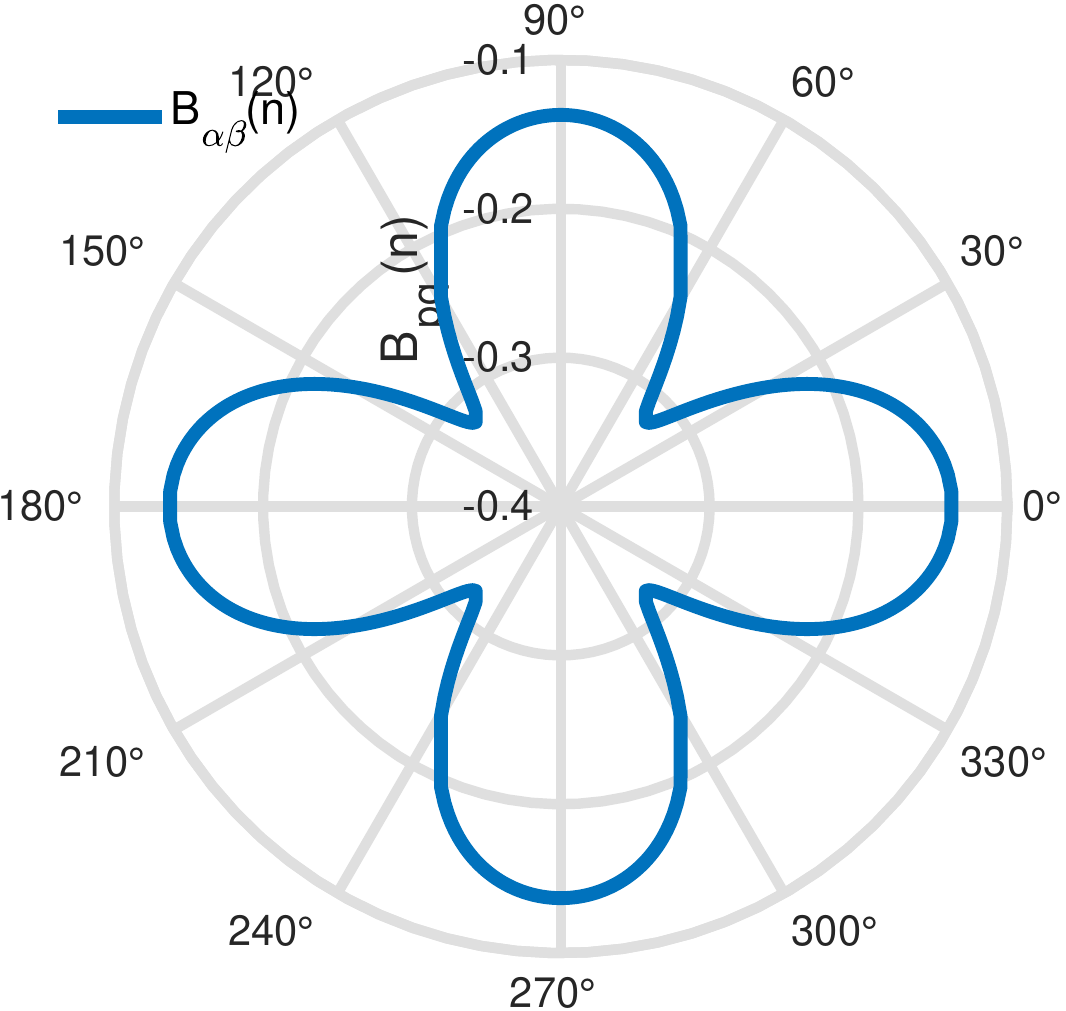}
  \caption{}
  \label{pol_y3}
\end{subfigure} 
\caption{Polar plots of elastic interaction energy
between $\alpha$ and $\beta$ domains $B_{\alpha\beta}(\textbf{n})$
for alloys (a)$Y_2$ and (b)$Y_3$.}
\label{fig:elinten_y123}
\end{figure}

In $A-$poor $Y$ alloys, $\alpha$ domains always align along elastically soft $\langle 10\rangle$
directions dictated by Zener anisotropy ratio ($A_z=3$).
In contrast, alignment and shape of $\beta$ and $\gamma$ domains
change with the variations in the misfit strains.
Since the sign and magnitude of misfits affect $B_{\alpha\beta}(\textbf{n})$
and $B_{\alpha\gamma}(\textbf{n})$, the inner envelope of orientation dependent 
elastic interaction energy can be used to predict the shape as well as the alignment 
of both $\beta$ and $\gamma$. $B_{\alpha\beta}(\textbf{n})$ and 
$B_{\alpha\gamma}(\textbf{n})$ shows similar behavior for these alloy systems.
Therefore, we only analyze the interaction energy between $\alpha$ and $\beta$ domains.
$B_{\alpha\beta}(\textbf{n})$ is zero for $Y_1$, which explains the absence of any 
orientational preference in $\beta$ domains of the alloy. 
Since $B_{\alpha\beta}(\textbf{n})$ has minimum along $\langle 10\rangle$ 
directions in $Y_2$ (Fig.~\ref{pol_y2}), the $\beta$ domains in the alloy 
are oriented along the same directions. Fig.~\ref{pol_y3} shows the polar plot of 
$B_{\alpha\beta}(\textbf{n})$ in $Y_3$ and it has a minimum along $\langle 11\rangle$  
directions. Hence the $\beta$ domains are arranged along $\langle 11\rangle$ 
directions in $Y_3$ alloy. $\gamma$ phase shows similar
orientational preference as $\beta$ in all these alloys. 

It is evident from our investigation that the morphological alignment
and shape of elastically coherent domains are strongly influenced by the interactions 
between anisotropy in the elastic moduli and the misfit strains
between the coexisting phases.
The elastic interaction energies between these
co-existing domains modify their shape and directional alignment based
on the elastically soft directions.
The morphological characteristics are also dependent on the alloy composition, 
since the kinetic paths of decomposition can affect the alignment of domains.

We investigated the effect of elastic anisotropy on spinodal decomposition in ternary alloys
by systematically varying the alloy composition $(c_B^0,c_C^0)$ and misfit strains
($\epsilon_{\alpha\beta},\epsilon_{\alpha\gamma}$). The variations in coherency strains
between the coexisting phases and alloy chemistry change the equilibrium compositions and  
phase separation sequence, thereby influencing the morphological attributes
of the domains. The interplay between elastic anisotropy and
relative misfits between the phases modifies the elastic interaction
energies between the domains depending on the sign and degree of misfit strains.
The inner envelopes of the polar plots of these elastic interaction energies match with 
the morphological patterns of the precipitates. Thus, these interactions determine the 
change in shape and directional alignment of the domains in the microstructure.
Therefore, we conclude that the preferential crystallographic directions of  
alignment of precipitates in three-phase alloys depend not only on the anisotropy
in elastic moduli, but also on the sign 
and magnitude of the misfit strains. 

\section*{Acknowledgements}
The authors are grateful for the financial support from R $\&$ D and SS, Tata Steel Limited.


\bibliography{thesis_ref}

\end{document}